\documentclass[11pt]{article}

\usepackage[pdftex]{graphicx}
\usepackage[T1]{fontenc}
\usepackage{hyperref}
\usepackage{color}
\usepackage{url}

\usepackage{amssymb}

\pdfoutput=1

\begin{document}

\title{Jets from Hollows}

\author{Thomas Seon, Elisabeth Ghabache, Arnaud Antkowiak \\
\\\vspace{6pt}
UPMC \& CNRS  (UMR 7190)\\
Institut Jean Le Rond d'Alembert\\
 4 Place Jussieu, F-75005 Paris, France}

\maketitle

\begin{abstract}
This fluid dynamics video presents the gravity-driven jets induced by the relaxation of 
large prolate bubbles or large holes at a free liquid surface.
It shows that the jets developing inside the bubble can be strong enough to give rise to 
liquid projections shooting out way above the free surface.
\end{abstract}

In this fluid dynamics video (\href{https://dl.dropbox.com/u/11494676/APSpackage/anc/SeonGoFM2012VideoHD.mpg}{high-quality} or  \href{https://dl.dropbox.com/u/11494676/APSpackage/anc/SeonGoFM2012VideoSD.mp4}{small size version}), we present a surprising violent jet dynamics following the relaxation of 
large prolate bubbles after bubble disconnection and large holes at a free liquid surface.

The experiment presenting jets in bubbles consists in a bubbling flow (releasing gas with constant 
flowrate from a submerged orifice) in a 
viscous liquid.
The liquid used is sugar cane syrup of viscosity 
$\mu=110$ mPa.s and surface tension $\gamma = 90$ mN.m$^{-1}$. 
Air injection is 
controlled by a mass flow meter (Alicat Scientific) that provides a constant 
flowrate. It allows us to achieve a wide range 
of airflows (from 0.01 to 10 $\ell$.min$^{-1}$) as we can see in the video.
The two last shots present the formation and the collapse of a large hole at a liquid surface.
This experiment consists in blowing air over a free liquid surface so as to form a depression. Upon the collapse of this depression, a gravity-driven jet develops.
The liquids used are water and a 
water-glycerol mixtures of viscosity $\mu = 800$ mPa.s and  surface 
tension $\gamma = 65$ mN.m$^{-1}$. 
The hollow and jet dynamics are 
recorded through ultra-fast imagery. To do so, the tank is back lit and images 
are obtained at $7000$ ($3000$ for the two last shots) frames per second using a digital high-speed camera 
(Photron SA-5). 

In our last paper \cite{Seon2012} we investigate these phenomena by carrying out 
experiments with various viscosities, surface tensions, densities and nozzle radii. We propose 
a universal scaling law for the jet velocity in both situations (bubble and hole), which unexpectedly involves the 
hollow height to the power 3/2. This anomalous exponent suggests an energy focusing phenomenon. We demonstrate experimentally that this focusing is purely gravity-driven.


\begin{thebibliography}{10}
\bibitem{Seon2012} T.~Seon, and A.~Antkowiak. \newblock Large bubble rupture sparks fast liquid jet. \newblock {\em Phys. Rev. Lett.}, 109 (014501), 2012.
\end{thebibliography}
\end{document}